\documentclass[review,12pt,authoryear]{elsarticle}

\usepackage{amssymb,amsmath,amsfonts,amsthm}
\usepackage{floatrow}
\usepackage{graphicx,caption,subcaption}
\usepackage{wrapfig}
\usepackage{ragged2e,threeparttable}
\usepackage{booktabs}
\usepackage{multirow}
\usepackage{subcaption}

\DeclareMathAlphabet{\mathbfit}{OML}{cmm}{b}{it}
\DeclareMathOperator*{\argmin}{arg\,min}
\newcommand{\CL}{\mathbfit{c}}
\newcommand{\cl}{c}

\newcommand{\dataset}{\mathcal{D}}

\newcommand{\pt}{\pi}
\newcommand{\sub}{S}
\newcommand{\Dist}{\mathbfit{D}}

\newcommand{\pamf}{\gamma}

\newcommand{\perm}{\boldsymbol{\sigma}}
\newcommand{\mass}{\alpha}
\newcommand{\discount}{\delta}
\newcommand{\temp}{\tau}

\newcommand{\comment}[1]{}

\begin{document}

\begin{frontmatter}

\title{Cluster Analysis via Random Partition Distributions}

\author{David B. Dahl \\ Jacob Andros \\ J. Brandon Carter}

\begin{abstract}

Hierarchical and k-medoids clustering are deterministic clustering algorithms
based on pairwise distances. Using these same pairwise distances, we propose a
novel stochastic clustering method based on random partition distributions.
We call our method CaviarPD, for \underline{c}luster \underline{a}nalysis
\underline{via} \underline{r}andom \underline{p}artition
\underline{d}istributions.  CaviarPD first samples clusterings from a random
partition distribution and then finds the best cluster estimate based on these
samples using algorithms to minimize an expected loss. We compare CaviarPD with
hierarchical and k-medoids clustering through eight case studies. Cluster estimates
based on our method are competitive with those of hierarchical and k-medoids
clustering. They also do not require the subjective choice of the linkage method
necessary for hierarchical clustering. Furthermore, our distribution-based
procedure provides an intuitive graphical representation to assess clustering
uncertainty.

\end{abstract}


\end{frontmatter}

\section{Introduction}
\label{intro}

Cluster analysis seeks to partition data into distinct subsets, or clusters,
such that observations in the same cluster are more similar than observations
from different clusters. There are numerous methods for cluster analysis with
applications in many fields \citep{dubes1976, jain1999}. As an unsupervised
learning technique, there is little consensus on how to validate the clustering
obtained. With many available clustering techniques, subjective choices must be
made which influence the outcome of a clustering procedure
\citep{islr}. It is also difficult to quantify the uncertainty associated with
clustering estimates.

Among the various algorithms for cluster analysis, agglomerative hierarchical
clustering remains one of the most used \citep{hca}. Agglomerative hierarchical
clustering is a heuristic method that takes as its input a matrix of pairwise
distances among items. Beginning with every item in its own cluster, the algorithm
sequentially merges the most similar clusters (based on the pairwise distances)
until all observations are in a single cluster. These nested clusters can be
visualized as a dendrogram. There are subjective decisions involved in
hierarchical clustering; namely, the linkage type used to build the tree
and the tree-cutting method used to obtain a clustering estimate.

Another commonly used class of clustering methods includes k-means and k-medoids,
which both seek to minimize the within-cluster distances from each point to a
point specified as its cluster center. k-means minimizes squared Euclidean
distances to each cluster centroid. In contrast, k-medoids  uses an actual
data point for each cluster center and thus can take any dissimilarity
measure as input \citep{kaufman1990}. The same input of a pairwise distance
matrix makes k-medoids a comparable procedure to hierarchical clustering. The most
influential choice a user must make in k-medoids clustering is the value of
$k$, that is, how many clusters the estimate should contain.

We propose CaviarPD: Cluster Analysis via Random Partition Distributions.  Like
hierarchical clustering and k-medoids, CaviarPD is based on pairwise distances, yet
it provides a unique way to assess clustering uncertainty.
The CaviarPD method relies on sampling from a random partition
distribution that is based on pairwise distances, e.g., the Ewens-Pitman
Attraction (EPA) distribution \citep{dahl2017}. Thus, like the other two methods,
the EPA distribution uses pairwise distances as input. Unlike hierarchical and
k-medoids clustering, this form of cluster analysis allows us to make probability
statements about clustering relationships, thereby quantifying the uncertainty of the
estimate. In doing so, CaviarPD provides an alternate way to visualize clusterings by
showing the pairwise probabilities that items are clustered together. An
implementation of the method is provided in the \texttt{caviarpd} package \citep{caviarpdlib} in
\texttt{R}, which is available on CRAN.

To compare hierarchical and k-medoids clustering with our proposed CaviarPD
method, we evaluate how well each method performs in eight different case
studies where the true partition of the data is known. We compare methods
using two partition loss functions, namely, Binder loss \citep{binder1978} and
VI loss \citep{meila2007,wade2018}. Through the case studies, we show the advantages of
CaviarPD over hierarchical and k-medoids clustering. All methods
tend to estimate the true partition of the data well; however, the hierarchical
clustering results are highly varied between linkages and choices of cutting the
tree. There is little statistical reasoning to guide the choice of linkage and
tree cutting. Furthermore, because CaviarPD gives the probabilities that
items are clustered together, CaviarPD provides additional
information about the clustering relationships beyond what
hierarchical or k-medoids clustering provide.

\section{Existing Distance-Based Clustering Methods}

\subsection{Clustering Concepts and Terminology}
We introduce common concepts used in both traditional clustering methods and
CaviarPD.
For a more thorough description of current clustering practices, we
suggest \citet{hca}. A clustering $\CL = (\cl_1, ..., \cl_n)$ gives labels for $n$
items in which items $i$ and $j$ are in the same cluster if and only if $\cl_i =
\cl_j$. Equivalently, a partition $\pi = \{ S_1, \ldots, S_q \}$ of integers $1,
\ldots, n$ is composed of mutually exclusive, non-empty, and exhaustive subsets
such that $i, j \in S$ implies that $c_i = c_j$.  We use the terms `clustering'
and `partition' interchangeably and note that the term `cluster' is synonymous
with `subset'.

In cluster analysis, we seek to cluster $n$ observations of a dataset $\dataset
= \{\mathbf{x}_1 ,\dots, \mathbf{x}_n\}$ into distinct groups so that
observations within a group are more similar than observations from different
groups. CaviarPD, like the other two methods described, relies on distance
information between observations in order to partition items into subsets.
The pairwise distances between observations $\mathbf{x}_i$ and $\mathbf{x}_j$
are calculated from a specified distance function $d(\mathbf{x}_i,\mathbf{x}_j)$.
One common distance function is Euclidean distance:
$d(\mathbf{x}_i,\mathbf{x}_j)=\sqrt{(\mathbf{x}_i -
\mathbf{x}_j)'(\mathbf{x}_i-\mathbf{x}_j)}$. The pairwise distances between
all items can be stored in an $n \times n$ distance matrix $\Dist$, where
$d_{ij}=d(\mathbf{x}_i,\mathbf{x}_j)$. The choice of distance metric is
also an important consideration in the analysis, but our task here is to
compare distance-based clustering methods given the user's chosen distance
matrix.

\subsection{Hierarchical Clustering}

In this section we highlight the decisions a user must make with hierarchical
clustering.  For a thorough introduction to hierarchical clustering, see
\citet{mma} or \citet{islr}.

Recall that in agglomerative hierarchical clustering, each observation begins
in its own cluster and the most similar clusters are sequentially merged until
all data points are in a single cluster. The criteria used to define similarity
between clusters is called the linkage and is computed from the pairwise
distance between items in each cluster. Hierarchical clustering requires that
a user decide which linkage to use and how to cut the dendrogram to obtain a
partition estimate. For a more detailed explanation of the
subset distance computations for each linkage type, see \citet{nielsen2016}.

\begin{figure}[t]
\centering
\includegraphics[width=0.49\textwidth]{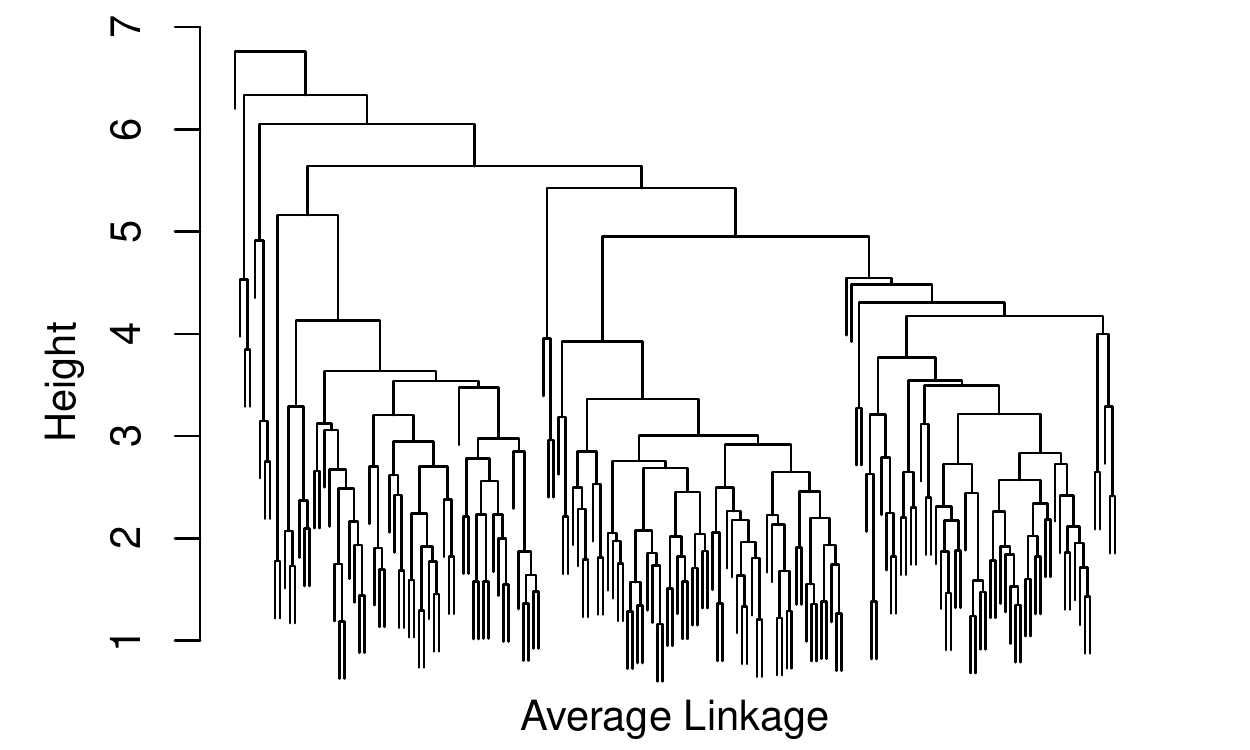}
\hfill
\includegraphics[width=0.49\textwidth]{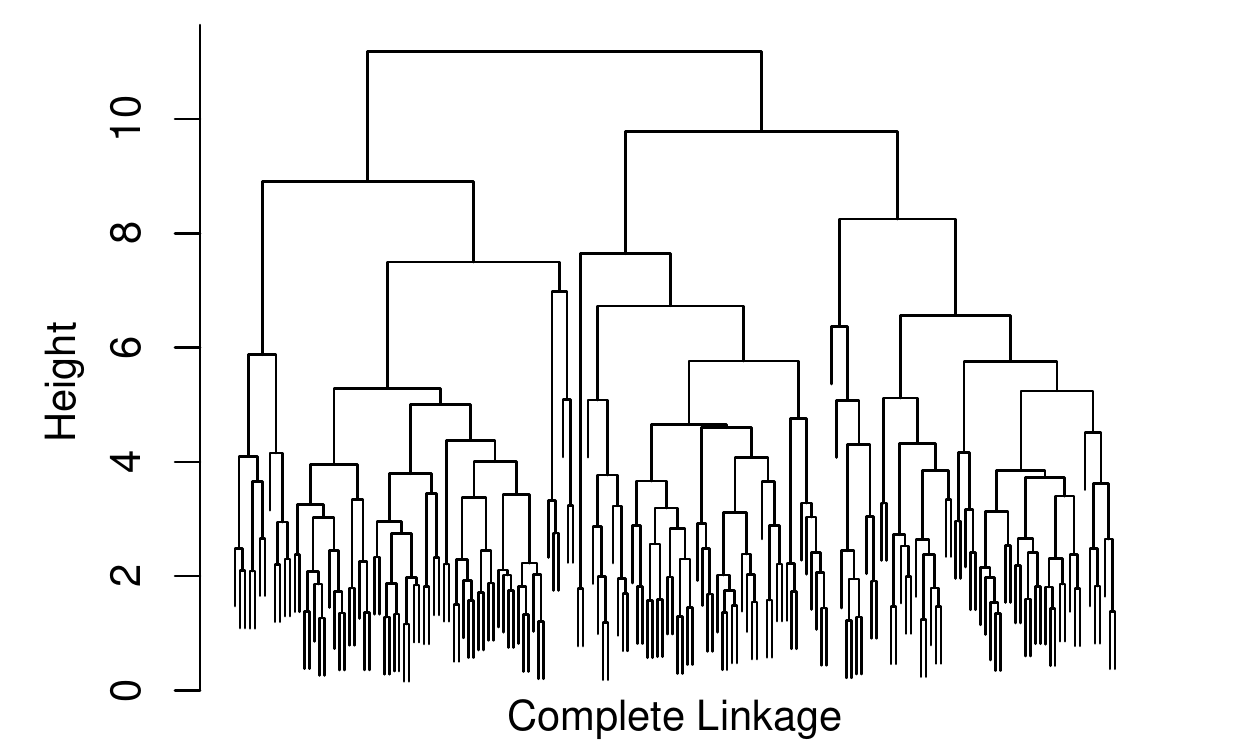}\\
\includegraphics[width=0.49\textwidth]{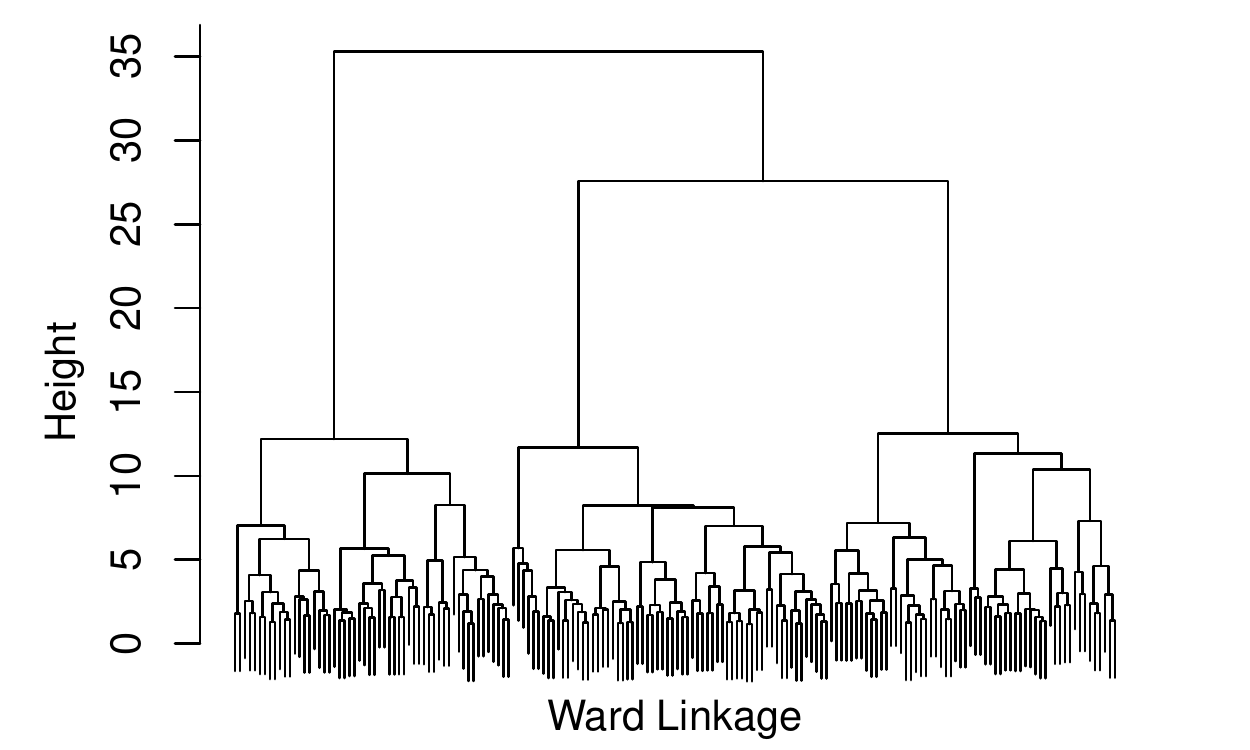}
\hfill
\includegraphics[width=0.49\textwidth]{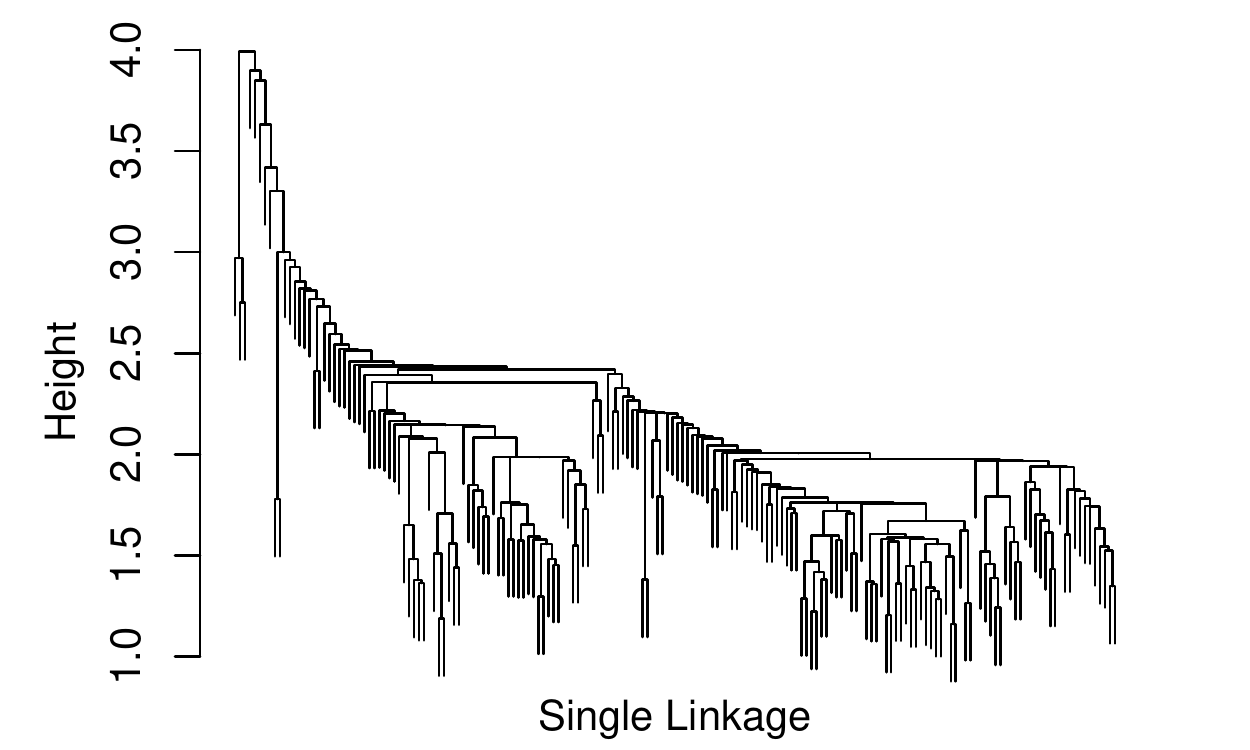}
\caption{Clustering dendrograms for the wine dataset using complete, average, Ward and single linkage.}
\label{fig:wine-linkage}
\end{figure}

We demonstrate how the choice of linkage leads to highly varied clustering
estimates. Figure~\ref{fig:wine-linkage} shows the resulting dendrograms
from four common linkages applied to the wine recognition dataset using the
function \texttt{hclust} in \textsf{R} \citep{r2018}. The wine dataset
contains many different chemical attributes for samples of wine from three
different cultivars. The distance matrix was computed with Euclidean distance
using all 13 chemical attributes. The clustering structure represented in the
dendrograms is very different for all four linkages. Single linkage produces
long chains of connected clusters. Ward linkage seeks to create compact
spherical clusters such that the dendrogram shows more distinct clusters
\citep{ward1963}. For the wine dataset, the dendrograms for average and complete
linkage show clustering structures between the long chains of single linkage and compact clusters of Ward linkage.

After choosing the type of linkage, the structure of the dendrogram is
used to cut the tree to obtain a clustering estimate. From the dendrogram
using Ward linkage, there appear to be three main clusters. Inspection of
the other dendrograms leads to less definite conclusions about the number
of clusters in the dataset as each tree varies drastically. Using complete
linkage, one could reasonably argue for a cut of the tree that gives
anywhere from 2 to 7 clusters. \citet{langfelder2008} introduce a solution
to the tree cutting problem with the dynamic tree cut (DTC) algorithm.
This procedure takes into account the structure of the tree to
detect clusters and also allows for the tree to be cut at variable heights,
providing greater flexibility in clustering estimation.

The lack of agreement between the linkages and tree cutting estimates is
prominent and concerning. There appeared to be three main clusters for ward
linkage, yet the DTC produced a default estimate of 5 clusters. The DTC applied
to the complete linkage dendrogram resulted in an estimate with 3 clusters.
Likewise, DTC applied to the average linkage dendrogram also resulted in 3 main
clusters with a single observation in a fourth cluster by itself. Lastly, the
single linkage tree resulted in a clustering estimate with all but 3
observations in a single cluster.

\subsection{K-Medoids Clustering}

k-means clustering remains one of the most common and popular clustering approaches
used today. The closely related  k-medoids clustering, like k-means, seeks to
minimize the sum of distances between points in a cluster and its cluster center.
Unlike k-means, k-medoids uses actual data points (exemplars or medoids) for the
cluster centers. Thus, k-medoids only uses the pairwise distance matrix as input,
making it a comparable clustering approach to hierarchical clustering and CaviarPD.

Since there are $\binom{n}{k}$ combinations of medoids that can be tested to
minimize the total distance, finding the exact solution to the equation is
difficult. \citet{kaufman1990} proposed the PAM algorithm (partitioning
around medoids) to conduct a nonexhaustive but greedy search through the
combination space. The PAM algorithm was then modified by
\citet{schubert2019} to reduce computation time even further at the expense
of some accuracy. Since the datasets in our case studies are all relatively
small, we can easily use the original PAM algorithm for our k-medoids comparison.

In our case studies we use the silhouette method, a straightforward procedure
for selecting $k$ in k-means and k-medoids clustering \citep{rous1987}. The
silhouette method is most reliable when the choices for $k$ are narrowed down
to a reasonable range using some prior intuition about the data. As with
hierarchical clustering, the k-medoids approach offers no way to assess
uncertainty in the results. We use the \texttt{pam} function in the
\texttt{cluster} package \citep{cluster} since it has a built-in calculation for silhouette
width.

\section{Cluster Analysis Via Random Partition Distributions}
\label{caviarpd}

We propose the CaviarPD method, a novel approach to clustering based on random
partition distributions. A reference implementation is available on CRAN in the
\texttt{caviarpd} package \citep{caviarpdlib}.
CaviarPD is based on the Ewens-Pitman Attraction (EPA) distribution,
originally proposed as a prior distribution for partitions in a Bayesian
framework \citep{dahl2017}. In the proceeding subsections, we explain how the
EPA distribution can be used directly for cluster analysis along with the additional
components of CaviarPD. Since the EPA distribution is a probability distribution over
partitions, our simulation-based method provides as output probabilities of pairs of items being
clustered together.

\subsection{Ewens-Pitman Attraction Distribution}

In the EPA distribution, observations are sequentially allocated to subsets of a
partition with probability proportional to the attraction of the item to that
subset. We use $\perm = (\sigma_1, ..., \sigma_n)$ to denote the order of the
$n$ observations that are sequentially allocated into subsets in order to form a
partition. Thus $\sigma_t$ is the $t$th observation allocated in the partition.
This sequential allocation of items yields a sequence of partitions for each
step of the allocation. Let $\pt(\sigma_1,...,\sigma_{t-1})$ denote the
partition of the first $t-1$ observations at time $t-1$. At each time $t$, let
$K_t$ be the number of subsets in the partition $\pt(\sigma_1,...,\sigma_{t})$.
When $t = 1$, the first item is allocated to a new subset by itself with
probability 1 as shown in equation \ref{epa2}. For $t > 1$, $\sigma_t$ can
either be allocated to one of the existing $K_{t-1}$ subsets or it can be
allocated to a new subset in the partition. The probability mass function is
conveniently expressed as the product of increasingly conditional probabilities:

\begin{equation}
\label{epa1}
p(\pt_n|\mass,\lambda,\perm) =
\prod_{t=1}^{n} p_t(\mass,\lambda,\pt(\sigma_1,...,\sigma_{t-1}))
\end{equation}
\begin{multline}
\label{epa2}
p_t(\mass,\lambda,\pt(\sigma_1,...,\sigma_{t-1})) =
\mathrm{Pr}(\sigma_t \in A \,| \, \mass, \lambda,
\pt(\sigma_1,...,\sigma_{t-1})) \\ =
\begin{cases}
\frac{t - 1}{\mass+t-1}\cdot \frac{\sum_{\sigma_s \in
A}\lambda(\mathbf{x}_{\sigma_t},\mathbf{x}_{\sigma_s})}{\sum_{s=1}^{t-1}\lambda(\mathbf{x}_{\sigma_t},\mathbf{x}_{\sigma_s})}
&\text{for} \; \sub \in \pt(\sigma_1,...,\sigma_{t-1}) \\
\frac{\mass}{\mass + t -1} &\text{for} \; \sub \;
\text{being a new subset}
\end{cases}
\end{multline}

The probability that $\sigma_t$ is allocated into subset $\sub$ is a function of
the similarity function $\lambda$ and the mass parameter $\mass$.\footnote{
\citeauthor{dahl2017} also include a discount parameter
$\discount$ that further influences sampling from the EPA distribution;
For simplicity, we set $\discount=0$, yielding (\ref{epa2}).}
The similarity function $\lambda(\mathbf{x}_i,\mathbf{x}_j)$ gives
pairwise similarity between observations $\mathbf{x}_i$ and $\mathbf{x}_j$ for any
$\mathbf{x}_i,\mathbf{x}_j \in \dataset$. \citet{dahl2017} propose a general
class of similarity functions $\lambda(\mathbf{x}_i,\mathbf{x}_j)=f(d_{ij})$,
where $f$ is a non-increasing function of the pairwise distance $d_{ij}$ between
observations $\mathbf{x}_i$ and $\mathbf{x}_j$. $d_{ij}$ is the same distance
matrix $\Dist$ used as input for hierarchical and k-medoids clustering. Two common
similarity functions are reciprocal similarity $f(d) = d^{-\tau}$ and
exponential similarity $f(d) = \mathrm{exp}(-\tau d)$. The parameter $\tau \ge
0$ is the temperature, which has the effect of accentuating or dampening the
distance between items.

\subsection{Sampling from the EPA Distribution}

Sampling from the EPA distribution is a straight forward process
\citep{dahl2017}. Because items are allocated with probability proportional to
the similarity to items in an existing subset, similar items are more likely to
be clustered together in simulation. In order to sample a partition $\pt_n$ from
the EPA distribution, we begin with a permutation (some ordering) of the data
and fixed $\mass$, $\lambda$. The first item $\sigma_1$ is allocated to a subset
by itself with probability 1. The next item, $\sigma_2$ can either the assigned
to the subset with $\sigma_1$ or to a new subset by itself.  The probability of
each allocation is given in equation \ref{epa2}. Let $\sigma_2$ be randomly
assigned to the existing subset or new subset with the respective probabilities.
For each subsequent item, the item $\sigma_{t+1}$ is randomly assigned to an
existing subset or new subset with probability respective to being assigned to
$\sub_1, \sub_2, ..., \sub_{K}, \sub_{K+1}$.  Here $\sub_1, ... , \sub_{K}$
represent the existing subsets of the partition and $\sub_{K+1}$ is a new
subset. Continue to sequentially assign the items until a partition of the data
is obtained. The resulting partition is a single draw from the EPA distribution.
Sampling can be parallelized over many cores to simultaneously obtain many draws
from the EPA distribution.

The order by which the data are sampled affects the resulting probabilities of
obtaining particular partitions. To remove this dependence on sampling order,
randomize the order by which the items are allocated into subsets for each draw.
This has the effect of making the probability of each partition independent of
any particular permutation of the data.

\subsection{Visualizing Pairwise Probabilities}

A key advantage of CaviarPD over traditional clustering is its ability to
quantify and visualize uncertainty in clustering estimates. This is done using a
heat map from a summary of the samples from the EPA distribution.

Each of the partitions $\pt$ can be represented as an $n \times n$ association
matrix denoted $\pamf(\pt)$, where the $(i,j)$ element of the association matrix
is an indicator that observations $\mathbf{x}_i$ and $\mathbf{x}_j$ are in the
same cluster. In short, $ \pamf_{ij}(\pt) = \mathrm{I}(\cl_i = \cl_j) $. For $B$
samples from the EPA distribution, there are $B$ $\pamf(\pt)$ matrices. These
matrices $\pamf(\pt_1),...,\pamf(\pt_B)$ can then be averaged together
element-wise to create a pairwise similarity matrix, which contains the
estimated pairwise probabilities that items appear in the same subset for a
given $\mass$ and $\lambda$. The pairwise similarity matrix is an $n \times n$
matrix denoted $\Psi$, where the $i$th, $j$th element is the relative frequency
with which items $i$ and $j$ are clustered together among the samples.
$$\mathrm{Pr}(\cl_i = \cl_j) \approx \Psi_{ij} =  \frac{1}{B} \sum_{k=1}^{B} \pamf_{ij}(\pt_k)$$
Each element of the $\Psi$ matrix is the estimated probability that observations
$\mathbf{x}_i$ and $\mathbf{x}_j$ are in the same cluster for a given $\lambda$
and $\mass$. $\Psi$ can then be conveniently visualized in the form of a heat map.

\begin{figure}[tb]
\centering
\includegraphics[width=2.5in]{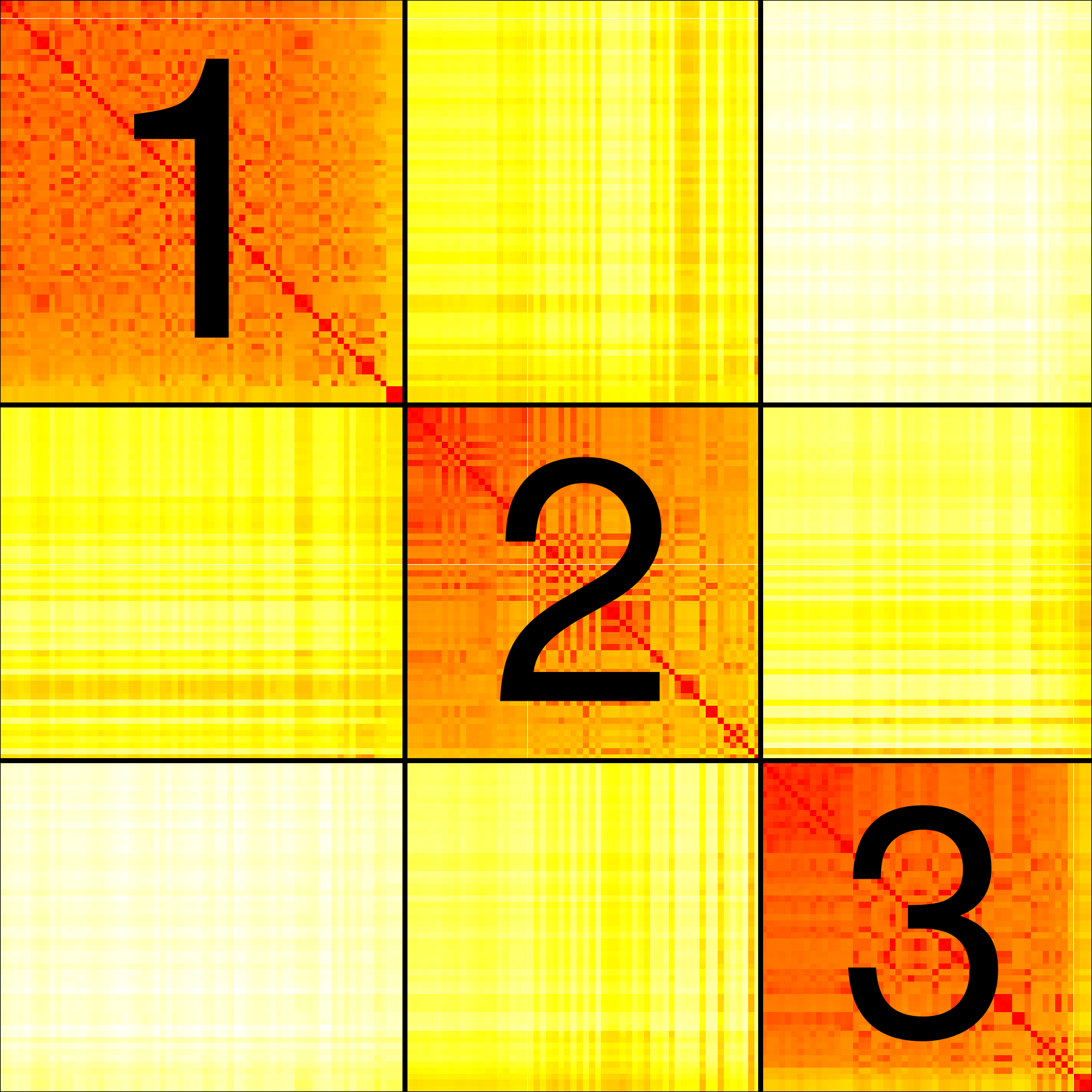}
\caption{Heat map of the pairwise probabilities for the wine dataset with $\mass=0.90$, $\discount=0$, and $\tau=10$.}
\label{fig:wine-conf-3clust_1}
\end{figure}

In the heat map, the color of a cell represents the probability that two
items are clustered together while also highlighting the actual clustering
estimate. In Figure~\ref{fig:wine-conf-3clust_1} we can see clearly that, for
$\mass=0.90$, there appear to be 3 distinct clusters. In this visualization, the
observations are ordered to group similar items together, creating the block
diagonal. The heat map becomes even more useful when the observations are
ordered by an estimated partition, thereby showing the probability relationships
within and between clusters. For example, clusters 1 and 2 appear much more
similar than clusters 1 and 3 because there is a darker shade of yellow in the
off-diagonal blocks for clusters 1 and 2, and a near-white shade of yellow in
the off-diagonals for clusters 1 and 3.

\subsection{Partition Estimation From Samples}

The Bayesian literature provides multiple methods for obtaining a point estimate
of a random partition based on samples from a posterior partition distribution
\citep{dahlsalsopaper}. Again, our paper here is not Bayesian, but we can draw
upon the Bayesian literature on random partitions to obtain a representative point
estimate of a partition based on samples from the EPA distribution.

In decision theory, a loss function is specified in order to pick an optimal
estimate that incurs minimal expected loss.  For partitions, loss functions
evaluate how distant the estimated partition is from the true partition of the
data.  Binder loss is a function of the number of disagreements
between the estimated and true partition for all possible pairs of observations
\citep{binder1978}.  The function is a weighted sum of the two types of
disagreements: observations are in different clusters when they should be in the
same cluster, and observations are in the same cluster when they should be in
different clusters. \citet{wade2018} demonstrate that when the weights of the
two errors are equal, $w_1=w_2$, then the partition that minimizes Binder loss
is given by:
$$ \hat{\pt}_{\text{binder}} = \argmin_{\pt \in \Pi} \sum_{i<j} (\pamf_{ij}(\pt) -
\Psi_{ij})^2.  $$
\citet{wade2018} also propose using the variation of information, introduced by
\citet{meila2007} as a loss function. The variation of information is developed
from information theory and is the information present in both clusters
(partition entropy) minus the information shared between the two clusters.
Minimization by enumeration over the whole partition space is
unfeasible, so we use the SALSO method \citep{dahlsalsopaper}
as implemented in the \textsf{R} package \texttt{salso}
\citep{dahl_salso}.

\subsection{Selecting the Mass Parameter}

\begin{figure}[tb]
	\centering
	\begin{subfigure}{.48\textwidth}
		\centering
		\includegraphics[width=1.75in]{figures/wine-conf-3clust.pdf}
		\caption{$\mass=0.90$}
		\label{fig:wine-conf-3clust}
	\end{subfigure}
	\quad
	\begin{subfigure}{.48\textwidth}
		\centering
		\includegraphics[width=1.75in]{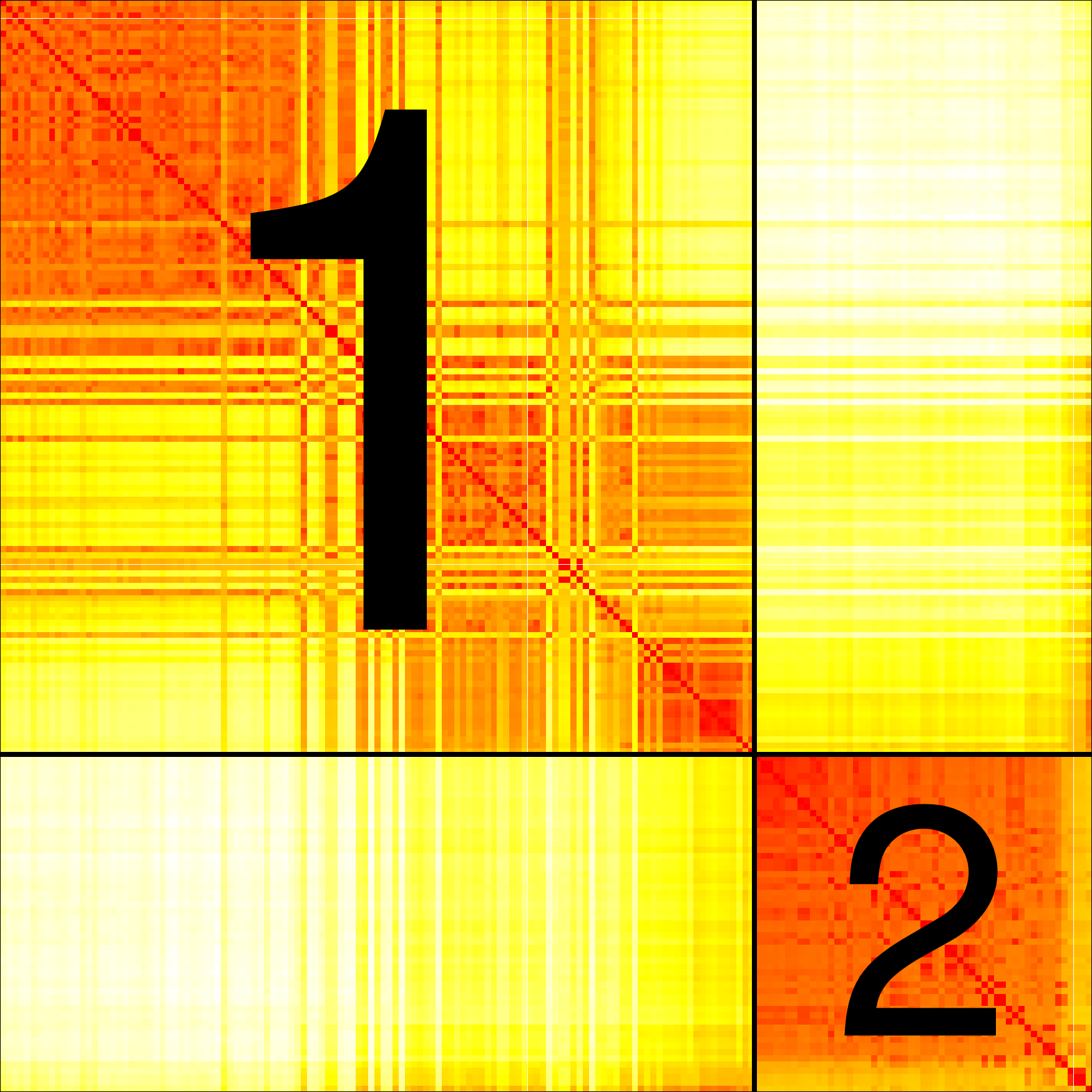}
		\caption{$\mass=0.70$}
		\label{fig:wine-conf-2clust}
	\end{subfigure}
	\caption{Heat maps of the wine dataset for two different mass parameters}
	\label{fig:wine-conf-plots}
\end{figure}

Most parameters in the EPA distribution have a default value. However,
the mass parameter $\alpha$ does not and it is highly influential in
determining the number of subsets in a partition estimate.
An objective algorithm is needed that can select a default mass
parameter for any given dataset. Ideally, the heat maps generated from
estimates with this algorithm should show that items in the same cluster
are clustered together with high probability. They should also show
clear distinction between subsets of the partition in the heat map (i.e.
low probability regions between clusters). Figure \ref{fig:wine-conf-plots}
shows two heat maps for the pairwise probabilities of the wine dataset
with the mass parameter set at 0.9 and 0.7, respectively. When the mass
is 0.9, the resulting heat map shows overall higher pairwise
probabilities within clusters as compared to the mass set at 0.7.
Likewise, there is less variance of the pairwise probabilities
within clusters when the mass is 0.9. The estimated three clusters
appear more distinct because of the greater within-cluster pairwise
probabilities. The goal of the mass selection algorithm, in the case
of the wine dataset, is to pinpoint $\alpha$ around 0.9 instead of 0.7.

We propose an objective mass selection procedure based on the
silhouette method, which is also used to select $k$ in k-means and
k-medoids clustering \citep{rous1987}. The `silhouette width' of a
given observation in any clustering estimate (not just k-medoids) is a
measure of compactness. It is proportional to the difference between the
observation's average within-cluster distance and its average distance
to points in the nearest cluster. The average silhouette width is the mean
silhouette width for all points in a dataset; seeking to maximize the
average width is intended to result in more compact and distinct clusterings.
Thus, the average silhouette width is a numerical summary that corresponds well
to the properties outlined in the preceeding paragraph of a ``ideal'' heat map.
In our mass selection algorithm, the user proposes a range of cluster
counts to consider in estimation. Boundary mass values are obtained for
the minimum and maximum of this range, and a grid search is performed
between these boundary masses. The mass from the clustering estimate
that results in the maximum average silhouette width is then used
for $\alpha$.

\comment{
$$ \hat{\pi}_{\text{sil}} = \frac{1}{n} \sum_{i=1}^n \frac{b(i) - a(i)}{\max\{a(i), b(i)\}} $$

$$ a(i) = \frac{1}{C_i - 1} \sum_{j \in C_i, i \neq j} d(i,j) \hspace{.2cm} \text{and} \hspace{.2cm} b(i) = \min_{k \neq i} \frac{1}{|C_k|} \sum_{j \in C_k} d(i,j) $$

}

\section{Case Studies}

To compare CaviarPD with the other distance-based clustering methods, we evaluate
how well each method clusters data where the true partition of the data is known.
Datasets were obtained from the UC Irvine machine learning repository
\citep{dua2017}. In order to evaluate the quality of the estimate we use
Binder and VI loss, which both measure the similarity between two
partitions. Both of these metrics are bounded to be nonnegative, where lower values
are indicative of a more accurate partition. We compare average, complete, and Ward
linkages to CaviarPD. Because the possibilities for tree cutting are numerous,
we cut the dendrograms using the default settings for the \texttt{cutreeDynamic}
function in the \texttt{dynamicTreeCut} package in \textsf{R} to obtain a
partition estimate for each of the linkages \citep{langfelder2016r}.

For each of the datasets with numeric attributes, we centered and scaled
the data and computed the Euclidean pairwise distances. For
datasets with categorical attributes, we computed the Jaccard pairwise
distances. Centering and scaling the data coerces the mean and standard
deviation of each predictor to be 0 and 1, respectively. Though not always
necessary, this helps safeguard against attributes with large variance carrying
more weight in the distance computation. We use all three linkage types from
hierarchical clustering to give it the best chance at competing with CaviarPD.
We also choose $k$ in k-medoids using the silhouette method. Despite these
advantages given to these other methods, CaviarPD still remains highly
competitive.

In CaviarPD, we fixed the temperature $\temp$ at 10 for the exponential
similarity function $\lambda$. These defaults seem to be sufficient for the
majority of applications of CaviarPD. We used our mass selection
algorithm to obtain a default value for the mass $\mass$ and corresponding
partition estimate. When using the mass selection algorithm, Binder loss and
VI gave very similar clustering estimates. For this reason, and for the sake of
simplicity, we use only the Binder loss CaviarPD estimates in our comparisons.
Again, we use the \texttt{caviarpd} function from the \texttt{caviarpd} package
in \texttt{R} to carry out the analysis.

\subsection{Wine Dataset}

For the wine dataset, CaviarPD estimates the true partition on par with the main
combinations of tree cutting and linkages from hierarchical clustering. However,
the partition estimates from hierarchical clustering are inconsistent in
their results. Without knowing the true partition of the data, we would not know
that average linkage produces the best estimate.

\begin{table}[tb]
\begin{tabular}{lrrrrrrrrrrr}
  \toprule[1.5pt]
  & \multicolumn{3}{c}{Wine} && \multicolumn{3}{c}{House Votes}  \\
  \cmidrule{2-4} \cmidrule{6-8}
  &                         K & Binder        & VI            & &  K  & Binder        & VI  \\
  \midrule
  CaviarPD &                3 & \textbf{0.09} & \textbf{0.68} & &  2  & \textbf{0.22} & \textbf{0.99} \\
  Average: Default DTC &    4 & \textbf{0.09} & 0.69          & &  3  & \textbf{0.22} & 1.03 \\
  Complete: Default DTC &   3 & 0.19          & 1.21          & &  10 & 0.43          & 2.82 \\
  Ward: Default DTC &       5 & 0.17          & 1.23          & &  6  & 0.37          & 2.20 \\
  K-Medoids &               3 & 0.12          & \textbf{0.68} & &  2  & 0.23          & 1.10 \\
  \bottomrule[1.5pt]
\end{tabular}
\caption{Clustering results for the wine and house votes datasets}
\label{tab:wine-house}
\end{table}

Table~\ref{tab:wine-house} gives the clustering results for the wine dataset,
which was used for demonstration in previous sections and contains 13 chemical
attributes on wines from 3 different cultivars. The number of clusters for a
particular estimate is given by $K$. The DTC default gives roughly
the correct number of clusters for average and complete linkage (the fourth
cluster for average linkage is a singleton cluster), but not for ward linkage.
Overall, average linkage and CaviarPD perform the best when
compared to the true partition under Binder loss; however, when using a VI
comparison, CaviarPD and k-medoids slightly outperform all of the DTC cuts.
These differences are minimal, but the estimates for Ward and complete linkage
fall severely short of the estimates produced by CaviarPD and average linkage.
In any case, it is worth noting the lack of consistency in the resulting
estimates from hierarchical clustering, and the ability of CaviarPD to cluster
just as accurately as the best DTC cut and k-medoids.

\subsection{House Dataset}

The house votes dataset contains voting records for the 1984 House of
Representatives. The class attribute is party affiliation, Republican or
Democrat. Having only two clusters yet nearly 500 total observations, this
data provides a rigorous test for all clustering methods. Results are displayed
in Table \ref{tab:wine-house}. CaviarPD and average linkage produce estimates
with nearly identical loss metrics. However, note that CaviarPD also identifies
the correct number of clusters at 2 (along with k-medoids).

\begin{figure}[tb]
\centering
\includegraphics[width=0.75\textwidth]{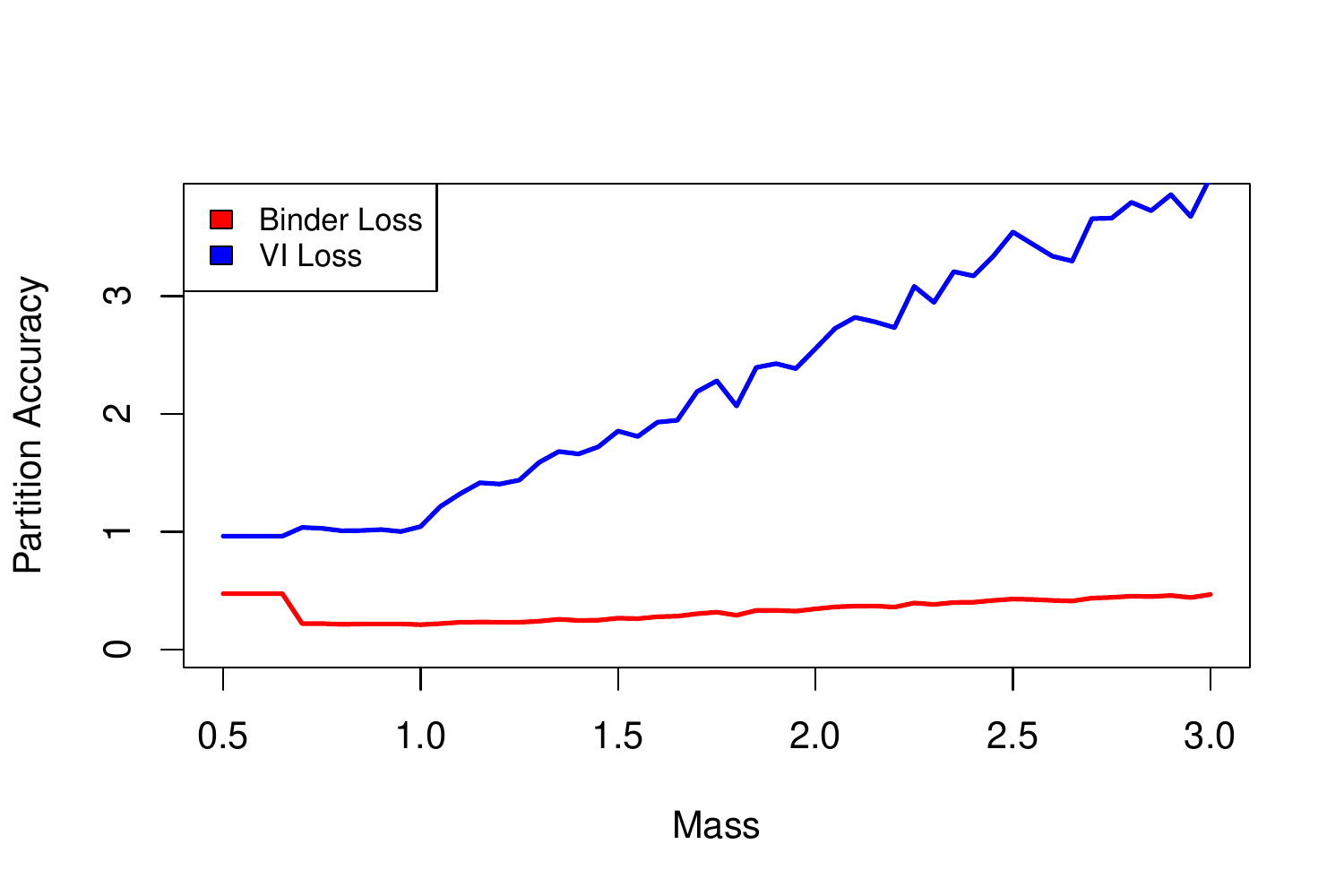}
\caption{Plot of Binder and VI loss across a grid of mass values for the house votes dataset.}
\label{fig:house-losses}
\end{figure}

The house dataset is also an excellent example for visualizing each loss
function as a function of the mass $\alpha$. Recall that the true partition is
only 2 clusters. Viewing Figure \ref{fig:house-losses}, note how the Binder
loss starts out high and immediately declines. This dip corresponds to the
clustering estimate moving from 1 cluster to 2 clusters as the mass increases.
Since the true partition is 2 clusters, this 2-cluster range of mass values
(between about $\alpha=.65$ and $\alpha=1.0$) results in the lowest Binder
loss. Moving up to the VI loss function, we see that VI loss penalizes less for
a small underestimation of the cluster count. As such, the 1-cluster estimate at
$\alpha=0.5$ is still deemed somewhat `accurate'. However, once the estimate moves
past the 2-cluster range, it quickly increases (particularly for VI loss),
making it clear that mass values any higher than 1.0 will not yield adequate
estimates. Thus, it becomes clear the importance of selecting viable mass
parameters.

\subsection{Flea-Beetle Dataset}

For smaller datasets, it is not uncommon to estimate the true partition of
the data. As an example, we take the flea-beetle dataset, which contains
measurements on three different species of beetles. It contains only 74
observations total. Both CaviarPD and Ward linkage (with the DTC cut) are
able to estimate the exact partition of the data. This results in the
comparative loss functions, both Binder and VI, being equal to 0.

\begin{table}
\begin{tabular}{lrrrrrrrrrrr}
  \toprule[1.5pt]
  & \multicolumn{3}{c}{Flea-Beetle} && \multicolumn{3}{c}{Olive}  \\
  \cmidrule{2-4} \cmidrule{6-8}
  &                         K & Binder       & VI           & &  K  & Binder        & VI \\
  \midrule
  CaviarPD &                3 & \textbf{0.0} & \textbf{0.0} & &  6  & 0.06          & \textbf{0.89} \\
  Average: Default DTC &    2 & 0.19         & 0.83         & &  10 & 0.07          & 1.05 \\
  Complete: Default DTC &   2 & 0.27         & 1.08         & &  14 & 0.12          & 1.69 \\
  Ward: Default DTC &       3 & \textbf{0.0} & \textbf{0.0} & &  9  & 0.11          & 1.25 \\
  K-Medoids &               3 & 0.02         & 0.17         & &  7  & \textbf{0.05} & 0.94 \\
  \bottomrule[1.5pt]
\end{tabular}
\caption{Clustering results for the flea and olive datasets}
\label{tab:flea-olive}
\end{table}

\subsection{Olive Dataset}

With the olive dataset we demonstrate the use of the pairwise similarity matrix
to detect clusters within subsets of the estimated partition.  The olive dataset
contains measurements on the levels of different oils in olives from nine
different regions of Italy.

The CaviarPD estimate, despite having the lowest VI loss, has
difficulty separating some pairs of regions, resulting in an estimate with
only 6 clusters. k-medoids encounters a similar problem and only detects
7 clusters. On the other hand, the average-linkage DTC estimate
concentrates far too many observations in the first 3 clusters, while also
creating a 10th cluster with only a few observations. The only hierarchical
clustering linkage that correctly estimates 9 clusters is Ward linkage;
however, Ward linkage misclassifies too many observations to be considered
as accurate as the other estimates.

\begin{figure}
	\centering
	\begin{subfigure}{.48\textwidth}
		\centering
		\includegraphics[width=1.75in]{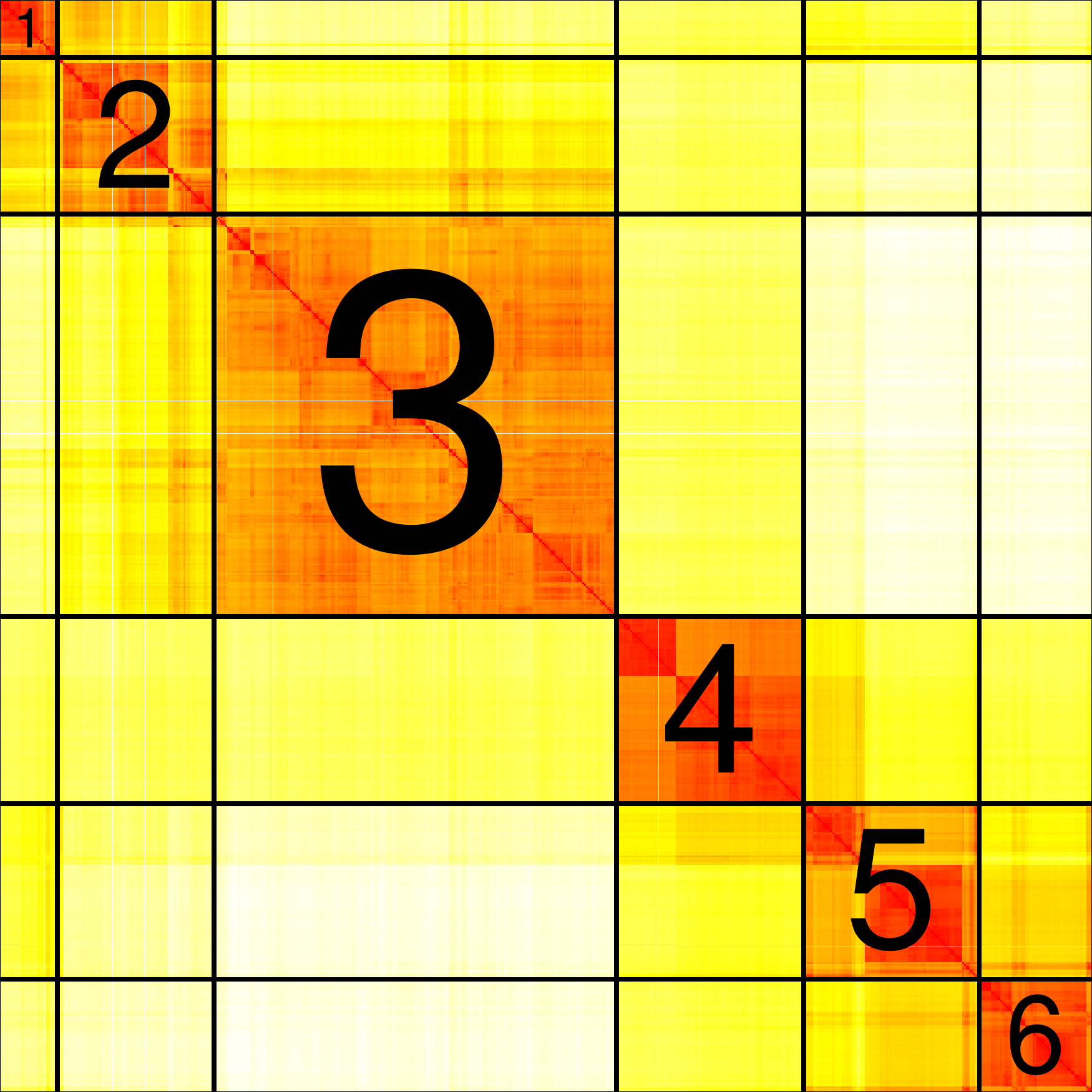}
		\caption{Olive dataset, $\mass = 2$.}
		\label{fig:olive-conf}
	\end{subfigure}
	\quad
	\begin{subfigure}{.48\textwidth}
		\centering
		\includegraphics[width=1.75in]{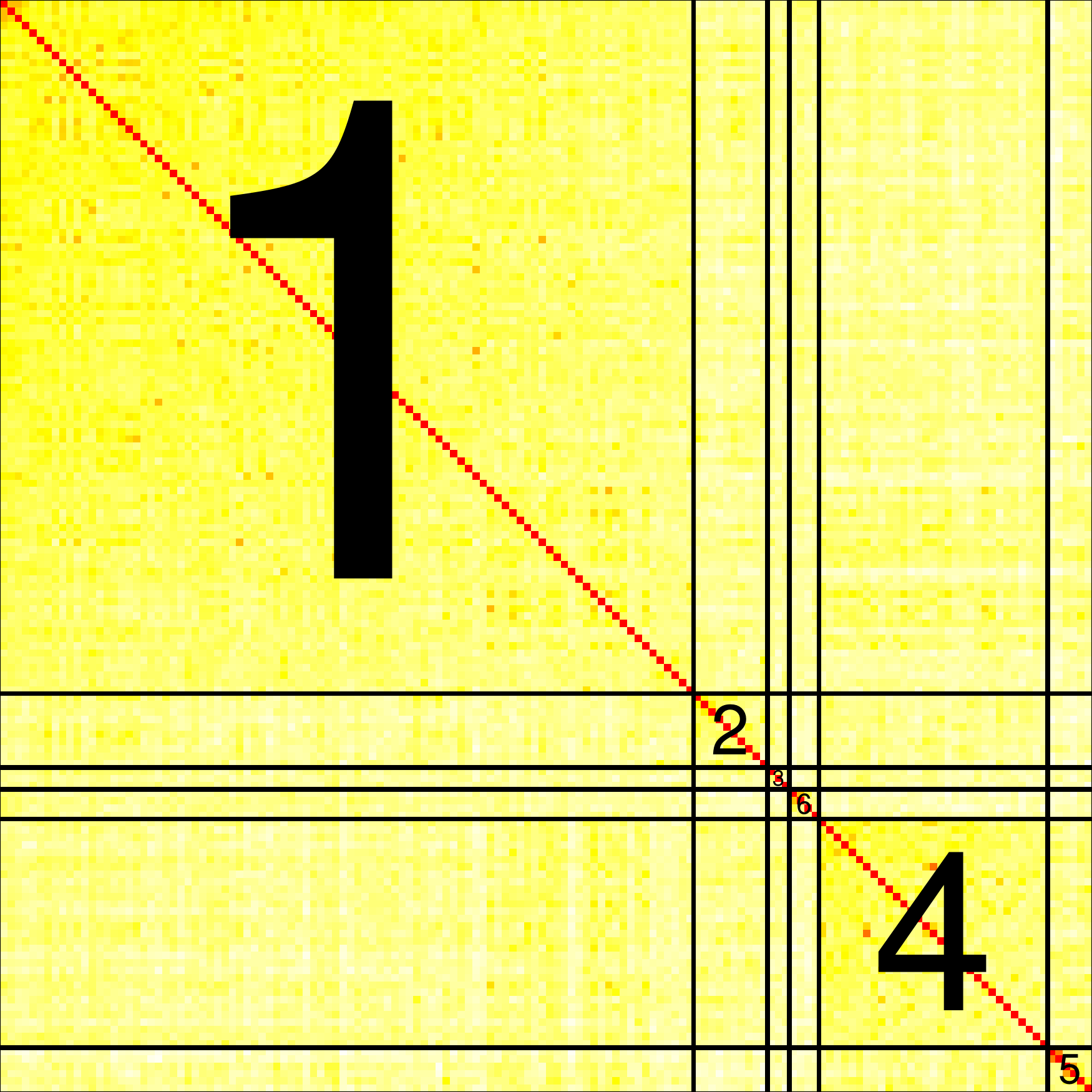}
		\caption{Lymphography dataset, $\mass = 1.1$.}
		\label{fig:lymph-conf}
	\end{subfigure}
	\label{olive-lymph}
	\caption{Heat maps of the olive and lymphography datasets}
\end{figure}

Heat maps from CaviarPD can be used to examine subcluster structure,
as demonstrated by clusters 4 and 5 in Figure~\ref{fig:olive-conf}.
In cluster 4 of the heat map, there appear to be two higher probability
regions of items being grouped together within each cluster. The same
is observed in cluster 5. In short, those clusters merged olives from
two different regions. Of course, post partition processing is also
possible in hierarchical clustering; however, in hierarchical clustering,
these decisions have no basis on probabilities as they do in CaviarPD.

\subsection{Additional Datasets}

The datasets in Table~\ref{tab:results_all} further demonstrate the
effectiveness of CaviarPD for partition estimation. The yeast dataset, which
contains 10 subsets in the true partition, is an example of a larger dataset
(approximately 1,500 observations) that makes running the entire mass selection
algorithm computationally intensive. Though the algorithm is still feasible
for a dataset this size, it may not be practical for significantly larger
sets. In such cases, we recommend building heat maps for 3 to 5
different values of $\alpha$ and selecting the mass for the plot with the
most concentrated pairwise probabilities. The lymphography
dataset is an example of data for which the 4 cluster distinctions are not
well-represented by the attributes. This leads to poor estimates by all
three methods, and a relatively uninterpretable
heat map as given in Figure \ref{fig:lymph-conf}. The E. coli dataset
contains attributes for 8 different localization sites of proteins in E. coli
bacteria. Chemical composition measurements are taken on 6 different glass
products in the glass dataset.

\begin{table}[tb]
\begin{tabular}{lrrrrrrrrrrr}
  \toprule[1.5pt]
  & \multicolumn{3}{c}{Yeast} && \multicolumn{3}{c}{Lymphography}  \\
  \cmidrule{2-4} \cmidrule{6-8}
  &                         K  & Binder        & VI            & &  K & Binder        & VI \\
  \midrule
  CaviarPD &                12 & 0.32          & 3.06          & &  6 & 0.49          & \textbf{1.44} \\
  Average: Default DTC &    5  & 0.51          & \textbf{3.00} & &  4 & \textbf{0.41} & 2.25 \\
  Complete: Default DTC &   19 & \textbf{0.24} & 4.83          & &  4 & 0.44          & 2.53 \\
  Ward: Default DTC &       16 & \textbf{0.24} & 4.70          & &  4 & \textbf{0.41} & 2.56 \\
  K-Medoids &               8  & 0.26          & 3.48          & &  2 & 0.43          & 1.96 \\
  \midrule
  & \multicolumn{3}{c}{E. Coli} && \multicolumn{3}{c}{Glass Products}  \\
  \cmidrule{2-4} \cmidrule{6-8}
  &                         K & Binder        & VI            & &  K & Binder        & VI \\
  \midrule
  CaviarPD &                6 & 0.12          & \textbf{1.22} & &  6 & 0.35          & 2.92 \\
  Average: Default DTC &    2 & 0.72          & 2.17          & &  3 & 0.50          & \textbf{1.90} \\
  Complete: Default DTC &   6 & \textbf{0.11} & 1.25          & &  4 & 0.42          & 2.42 \\
  Ward: Default DTC &       6 & 0.18          & 1.65          & &  4 & 0.35          & 3.05 \\
  K-Medoids &               6 & 0.17          & 1.80          & &  6 & \textbf{0.34} & 2.93 \\
  \bottomrule[1.5pt]
\end{tabular}
\caption{Clustering results for additional datasets}
\label{tab:results_all}
\end{table}

\subsection{Discussion}

We do not claim that CaviarPD
vastly outperforms all hierarchical clustering and k-medoids methods, as
that was not usually the case in these eight case studies. Rather, we have
shown that CaviarPD performs comparably to (and in some cases, slightly
better than) the other methods.  The benefit of CaviarPD, however, is the
ability to assess clustering uncertainty in the
form of the heat map plots from the pairwise similarity matrix.

In hierarchical clustering, there is no one linkage type or cutting technique
that consistently produces the best cut of the tree. We compared the best
possible DTC cut of the dendrogram for each linkage, but there is no cutting
rule that will consistently guide the user to that result. In k-medoids
clustering, there is less variability, but still no way to express uncertainty.
In contrast to these techniques, CaviarPD has a validating method to select
the mass parameter $\mass$. It also produces a pairwise similarity matrix which
provides a clear and consistent representation of how the data are grouped.

\subsection{Other Distributions}

The EPA distribution is suited for the CaviarPD method since it is a random
partition distribution taking pairwise distances as input.  Most other
random partition distrtibutions in the Bayesian literature do not take pairwise
probabilities and would therefore not be suitable for CaviarPD.  There is, however,
one other partition distribution which does, called the distance-dependent Chinese Restaurant
Process (ddCRP) \citep{blei2011}. The \texttt{caviarpd} function
allows users to specify the ddCRP distribution rather
than the default EPA distribution. Hence, clustering with the ddCRP distribution
from a coding standpoint is merely one additional argument.
However, default parameters for clustering with the ddCRP distribution have not
been investigated, and we have not found a good method for selecting optimal mass
for the ddCRP.

\begin{figure}
	\centering
	\begin{subfigure}{.48\textwidth}
		\centering
		\includegraphics[width=1.75in]{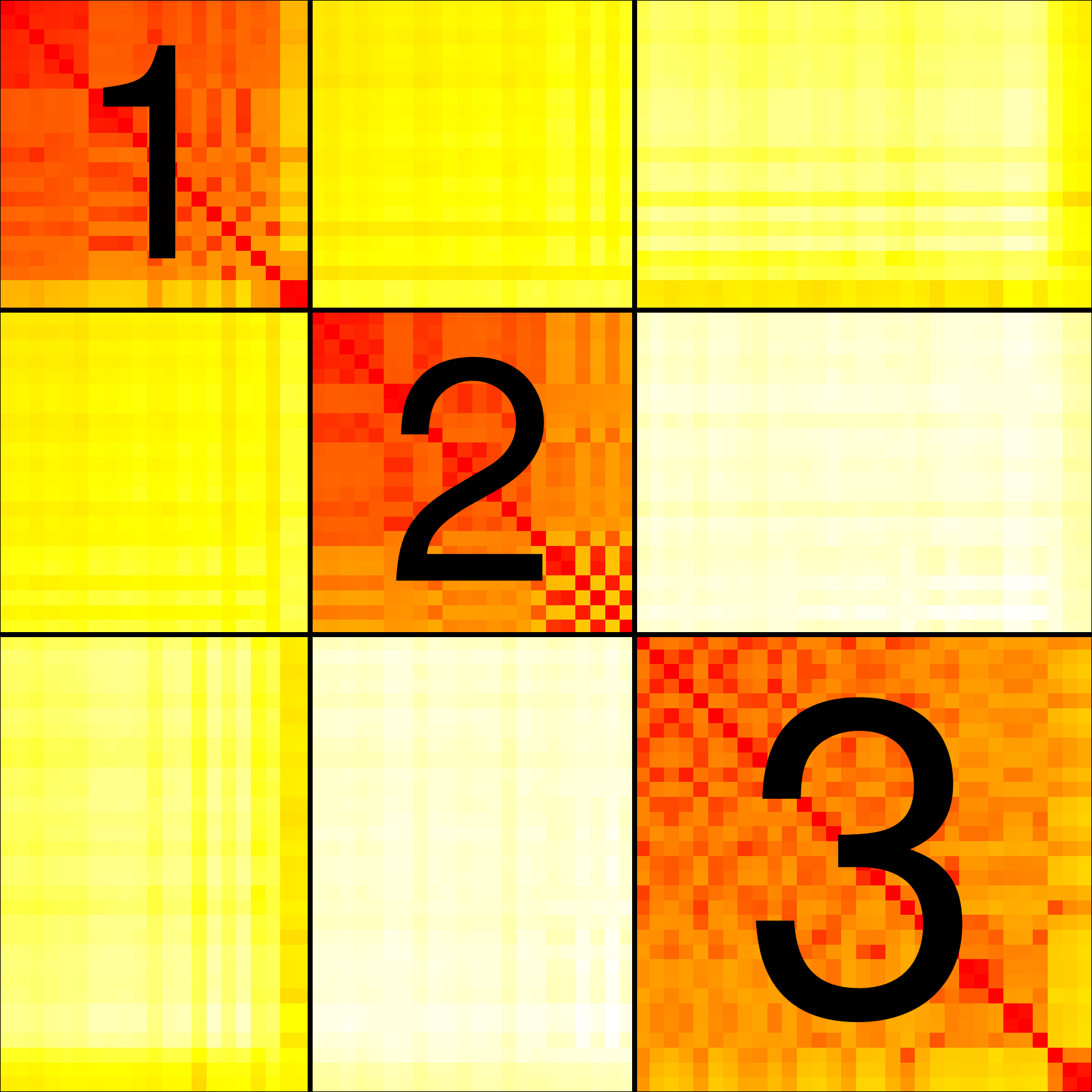}
		\caption{EPA Distribution}
		\label{fig:flea-epa}
	\end{subfigure}
	\quad
	\begin{subfigure}{.48\textwidth}
		\centering
		\includegraphics[width=1.75in]{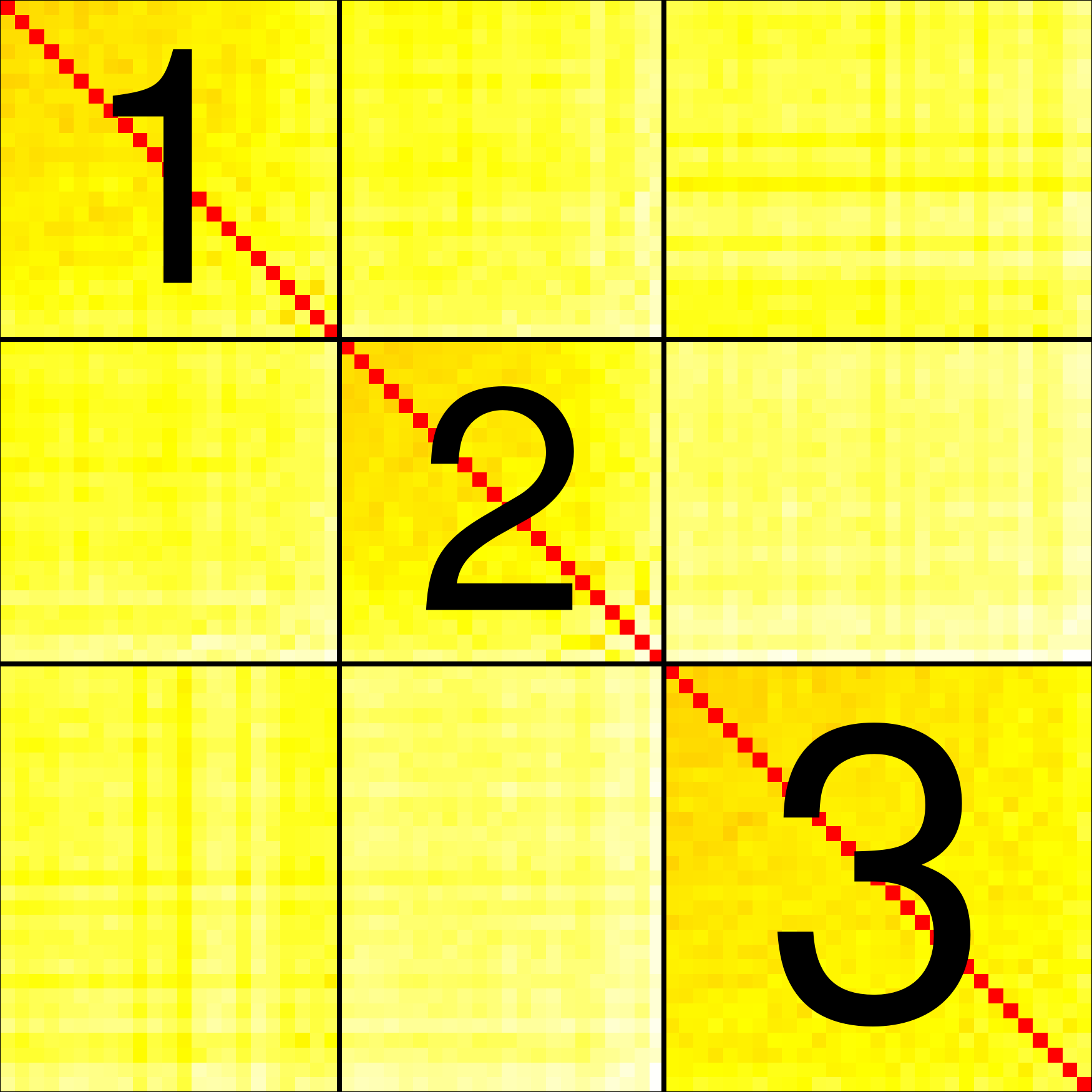}
		\caption{ddCRP Distribution}
		\label{fig:flea-ddcrp}
	\end{subfigure}
	\caption{Heat maps of the flea-beetle dataset. The EPA clustering estimates the exact partition, while the ddCRP clustering estimates a near-exact partition.}
	\label{fig:flea-conf-plots}
\end{figure}

In addition, while the ddCRP distribution does sometimes generate estimates that
are comparable with CaviarPD, it does not appear to lead to effective visualizations.
As an example, we take the flea-beetle dataset in Figure
\ref{fig:flea-conf-plots}, for which CaviarPD (and Ward linkage) were able to
estimate precisely the true partition. Plotting the estimate from the EPA
distribution shows clear distinction between the 3 clusters. All items in the
same cluster have high probabilities of being clustered together, and we can see
that elements in clusters 2 and 3 are extremely unlikely to be grouped together.
The plot from the ddCRP estimate shows none of this - it is virtually monochromatic
(besides the diagonal 1.0 probabilities of each item being clustered with
itself).

Estimates with the ddCRP distribution do not seem to be any more effective than
those from the EPA. They also are unable to provide the key insights that the
EPA estimates do, and have far more subjectivity in tuning the parameters. For
these reasons, we do not currently recommend using the ddCRP for the CaviarPD method.

\section{Conclusion} \label{conclusion}

Cluster analysis via random partition distributions simplifies the user's
dilemma for clustering. In hierarchical clustering, the different linkages and
lack of consistent tree cutting rules present a user with many subjective
choices. These different choices in hierarchical clustering lead to highly
varied partition estimates. For both hierarchical and k-medoids clustering,
there are no probability statements one can make regarding the clustering.
In contrast, data driven statistics help guide the
user in clustering estimation for CaviarPD. The pairwise similarity matrix
$\Psi$ gives a probabilistic understanding for how items are clustered together
in a dataset. Algorithmic clustering methods are not based on a
probability distribution and therefore quantify clustering uncertainty is difficult.

The central weakness of CaviarPD is the computational cost to select
a mass for large datasets. Hierarchical clustering, k-medoids (and the
corresponding k-means method) all have the potential to
run more quickly in these cases. It is also worth noting that while the
heat maps can be very insightful, they do not always visualize
overall clustering structure as comprehensively as a dendrogram would
in hierarchical clustering \citep{sander}. In summary, CaviarPD is not
the undisputed best clustering method in every instance. Rather, it
provides unique advantages over traditional approaches in the majority
of clustering problems.

\pagebreak
\bibliographystyle{elsarticle-harv} \bibliography{references}

\end{document}